# AN EVALUATION FRAMEWORK OF END-TO-END 5G MILLIMETER WAVE COMMUNICATION FOR CONNECTED VEHICLE APPLICATIONS

**Zadid Khan***
**Ph.D. Student**
Glenn Department of Civil Engineering, Clemson University
351 Flour Daniel, Clemson, SC 29634
Tel: (864) 359-7276; Fax: (864) 656-2670;
Email: mdzadik@clemson.edu

**Sakib Mahmud Khan, Ph.D.**
**Postdoctoral Scholar**
California Partners for Advanced Transportation Technology
Institute of Transportation Studies
University of California, Berkeley
Berkeley, CA 94720
Tel: (864) 569-1082
Email: sakibk@berkeley.edu

**Mashrur Chowdhury, Ph.D., P.E., F.ASCE**
**Eugene Douglas Mays Endowed Professor of Transportation**
Glenn Department of Civil Engineering, Clemson University
216 Lowry Hall, Clemson, South Carolina 29634
Tel: (864) 656-3313   Fax: (864) 656-2670
Email: mac@clemson.edu

**Mizanur Rahman, Ph.D.**
**Postdoctoral Fellow**
Glenn Department of Civil Engineering, Clemson University
200 Lowry Hall, Clemson, SC 29634
Tel: (864) 650-2926; Fax: (864) 656-2670;
Email: mdr@clemson.edu

**Mhafuzul Islam**
**Ph.D. Student**
Glenn Department of Civil Engineering, Clemson University
351 Flour Daniel, Clemson, SC 29634
Tel: (864) 986-5446; Fax: (864) 656-2670;
Email: mdmhafi@clemson.edu

*Corresponding author

Abstract: 239 words + Text: 6,367 + 1 tables: 250 = **6,856** words

Submission date: August 1st, 2020

**ABSTRACT**
The internet-of-things (IoT) environment connects different intelligent devices together and enables seamless data communication between the connected devices. Connected vehicles (CVs) are one of the primary example of the IoT, and the efficient, reliable, and safe operation of CVs demands a reliable wireless communication system, which can ensure high throughput and low communication latency. The 5G millimeter wave (5G mmWave) wireless communication network offers such benefits, which can be the enabler of CV applications, especially for dense urban areas with high number of CVs. In this study, we present a simulation-based evaluation framework of end-to-end 5G mmWave communication for CV applications. In addition, we compare the 5G mmWave with the Dedicated Short Range Communication (DSRC) technology for a CV application. The simulation framework is developed using two simulators, a network simulator and a traffic simulator. In order to develop the framework in this study, we have used Network Simulator 3 (ns-3) and SUMO, an open-source microscopic roadway traffic simulator. We have used end-to-end latency, packet loss and throughput as the performance evaluation metrics. We have found that for dense urban areas, 5G mmWave can achieve higher throughput, lower latency and lower data loss compared to DSRC. 5G mmWave can support CV applications with high throughput requirement on the downlink data flow. Through further investigation, we have found that the performance of 5G mmWave is significantly impacted by the penetration level of CVs, maximum CV speed, and CV application requirements.

**Keywords:** Connected vehicle, Automated vehicle, IoT, 5G, mmWave, DSRC

**INTRODUCTION**

Innovations in the domain of wireless communication technology have supported the massive growth of interconnected personal devices and intertwined sensory equipment owned by different agencies. However, the everlasting emergence of connected devices has increased the demand on the research community to develop a wireless communication enabler to support continuous and reliable data flow between the connected devices. Many are thinking that the fifth-generation wireless communication network or 5G will be the enabler to connect and support numerous connected devices under the broad umbrella of the internet-of-things or IoT *(1)*. IoT refers to the networked integration of highly distributed day-to-day devices via embedded systems *(2)*. 5G wireless communication, once deployed, has the potential to provide better communication reliability to the IoT interconnected devices compared to the existing fourth-generation (4G) and long-term evolution (LTE) wireless communication. Here it should be mentioned that, LTE is slower than "true 4G", as per the standards set by the International Telecommunications Union Radiocommunications Sector (ITU-R), but significantly faster than 3G, so LTE is used interchangeably with 4G *(3)*.

In transportation, connected vehicles (CVs) are one of the major components of IoT. Numerous safety, mobility and environmental benefits can be derived from CV applications *(4)*. As identified in previous literature, the advantages of 5G over 4G, LTE and LTE-V (LTE for vehicles) includes: increased spectrum allocation, availability of directional beamforming antennas used at both 5G enabled base stations and mobile entities, increased capacity to aggregate numerous simultaneous users within the coverage area, highly increased bit rates within increased proportions of the 5G coverage areas, and lower cost of infrastructure *(5)*. In order to increase communication network reliability, throughput, and connectivity in a IoT environment, researchers are showing more interest than before on spectrums higher than 6 GHz. The frequencies under 6 GHz is already divided into different LTE spectrums. The unused spectrums above 6 GHz have the potential to provide higher throughput for the increasing number of CVs in the future, especially in dense urban areas with a high number of CVs. The term 'mmWave' in '5G mmWave' simply refers to the spectrum with associated wavelengths in millimeters *(6)*. Compared to LTE, the carrier frequency of 5G mmWave allows for highly increased data rates while reducing the communication latency. Due to this inherent capacity offered by 5G mmWave for both backhaul links (within multiple base stations) and access links (within the base station and end users), it can support the CV environment where multiple data-intensive safety, mobility, and in-vehicle infotainment applications can run in numerous CVs.

The CVs can be equipped with in-vehicle devices containing dynamic beamforming-enabled multi-element antennas to connect with the mmWave base stations. The coverage area of the mmWave base stations are limited compared to LTE base stations. To overcome this issue, the base stations should be closely spaced. The wireless network operators are continuously reducing the cell coverage areas to enhance relaying, implement cooperative multiple-input-multiple-output (MIMO) antennas at the receiver and sender, and reduce inference *(5)*. As a result, the number of base stations will increase in the future mmWave based deployments, which will be applicable to the future CV environment where an unprecedented number of CVs will demand multi-gigabit/second to support CV applications and to stream real-time in-vehicle infotainment components. The applications of mmWave communication have been studied earlier. For the mobile communication system, Dehos et al. have identified mmWave as the primary technology for next-generation communication *(7)*. Mastrosimone and Panno have studied the performance of hybrid mmWave and LTE access links and compared the hybrid system against pure LTE-based access links, and they have found CVs can achieve increased throughput of 33% using the hybrid mmWave and LTE access links compared to the only LTE scenario *(8)*.

Recent pilot deployment sites in the US for connected vehicle applications are mostly equipped with Dedicated Short-range Communication (DSRC)-based devices, and in a few cases, a large area is wirelessly connected with combined DSRC and LTE to support the CV safety and mobility applications. Cellular network service providers are pursuing the plan to deploy a 5G network commercially. However, there is currently no evaluation framework for end-to-end performance of 5G mmWave communication for CV applications. Beside real-world tests, studies conducted in a simulation environment can provide extensive understandings of 5G mmWave benefits for different CV applications in different scenarios, such as different penetration of CVs, maximum CV speeds and CV application requirements. Several studies have investigated the 5G mmWave network for vehicular communication. However, the feasibility of the 5G mmWave has not been thoroughly studied. The objective of this research is to develop a simulation-based evaluation framework for the end-to-end 5G mmWave communication for CV applications and investigate the effect of CV penetration level, maximum CV speed, and CV application requirements on 5G mmWave performance using the framework.

**LITERATURE REVIEW**
This section includes literature review in the following two subsections: (i) 5G status in US and worldwide and (iii) 5G Application in Vehicular Networks.

**The Status of 5G**
In order to provide a higher data rate, the telecommunications industry and academia have been involved in research with the goal of improving the spectral communication efficiency by deploying more LTE base stations. Even with the significant enhancement in heterogeneous networks (HetNet) that include 4G, LTE, LTE-V, DSRC, and WiFi, these technologies are falling behind to meet user data requirements. The advent of the 5G era is expected to solve the increasing demand of data traffic. To date, LTE communication systems have been adopted in the US and worldwide. Every 10 years, a new generation of emerging communication technologies is replacing the old technologies since 1980: first generation analog FM cellular systems in 1981, second-generation (2G) digital technology in 1992, 3G in 2001, and LTE-A in 2011 *(3)*.

The vision of 5G is to offer a gigabits-per-second experience to end-users. The 3rd generation partnership project (3GPP) is combined of seven standards organizations and they are responsible for creating 5G new radio (NR) standards. They released the first full set of 5G NR standards in the release 15 in 2017 *(8)*. However, they are working towards updating the standards in the releases 16 and 17, which will cover vehicle-to-everything (V2X) application layer services, based on data collected from pilot real-world tests conducted in various countries *(9, 10)*. In the US, the telecommunication industry including AT&T, Verizon, and Sprint have already deployed 5G as a fully operational network in major cities such as Atlanta, Boston, New York, Chicago, San Francisco and Houston *(11)*.

**5G Application in Vehicular Networks**
The following issues pose challenges for the reliability of wireless communication for CV applications: (i) the dynamic network topology of vehicular communication changes very frequently due to high vehicle mobility and results in frequent data flow disconnections *(12)*, (ii) cross-channel interference in vehicle-to-vehicle communication increases packet drop rates when two adjacent channels are operated simultaneously *(13)*. Specifically, in a higher vehicle density scenario, the intensity of channel contention among vehicles increases significantly, which results in a higher transmission collision rate and a larger channel access delay *(14)*.

To overcome these challenges, 5G mmWave is expected to expand and support various application scenarios, such as enhanced Mobile Broadband (eMBB), Ultra Reliable Low Latency Communication (URLLC), and Massive Machine Type Communication (MMTC). eMBB is designed for the high data rate mobile broadband services, which require seamless data access both indoors and outdoors. In addition, URLLC is designed for applications that have stringent latency and reliability requirements in highly mobile vehicular communications to enable the CV network. MMTC supports a wide number of devices which sporadically generated small amount of data. Many recent studies have shown that 5G mmWave can be applicable for connected vehicles and V2X communications because of its high communication bandwidth with a gigabit/sec data rate and low latency communication delay *(15-18)*.

From the literature review, we have identified that there is a need for 5G mmWave evaluation framework for CV applications in a simulation environment. In this study, we aim to meet this research need by developing a simulation-based 5G mmWave evaluation framework for CV applications.

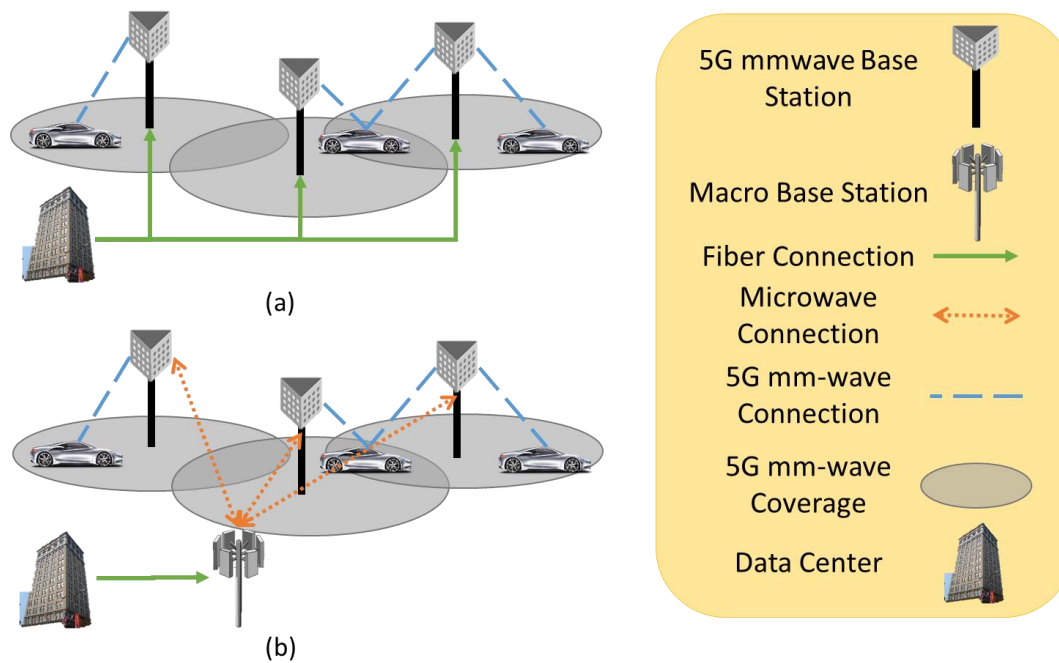

**FIGURE 1 5G mmWave network in urban environment**

**5G MMWAVE EVALUATION FRAMEWORK FOR CV APPLICATIONS**
In a connected environment, a data-intensive CV application is the in-vehicle infotainment system. Streaming high-quality video to a massive number of CV users demands a high data rate for wireless communication *(19)*. Automated vehicles (AVs) have different in-vehicle sensors, which generate a large amount of data that can be shared with other AVs and/or external computing infrastructure *(20)*. The 5G mmWave thus becomes a viable communication option for CVs and connected AVs. In the US, wireless communication providers like AT&T, Verizon, and T-Mobile have included 5G mmWave in their overall 5G communication deployment plan, which would potentially lead to the scenario of accessing the 5G mmWave network for CV applications in future. **FIGURE 1** shows the 5G mmWave enabled CV scenario for an urban area where a data center is connected with the mmWave base station either: (a) directly with the fiber optic network or (b) via macro base stations. The macro base station can be wirelessly connected with the data center with microwave connection for non-line-of-sight scenarios.

In this section, we present the 5G mmWave evaluation framework developed in this study. Figure 2 shows the high-level steps for developing a 5G mmWave evaluation framework for supporting CV applications. At first, we need to model the different OSI (Open System Interconnection) layers of the 5G mmWave communication. The lower layers, including the channel model, the PHY layer and the MAC layer has many unique characteristics in terms of 5G mmWave. After that, the upper layers are similar to LTE internet stack (TCP/IP layers). Then, we need to provide the CV application specifications, such as uplink/downlink data flow, data rate and packet size. After modeling the OSI layers, we need to provide the network mobility model for all moving nodes in the simulation. The 5G mmWave base station (i.e. evolved nodeB or eNB) and other communication infrastructure (i.e. remote host) will have a constant position mobility model. For CVs, we need to use a microscopic traffic simulator to generate the trace file containing information about the motion of CVs, such as CV position, CV speed and timestep when the CV data is captured. The trace file is used to generate the network mobility model, which simulates moving nodes in the network simulator. The network mobility model is used by the network simulation model and raw output files are generated. The output files are post-processed to generate evaluation metrics such as throughput, end-to-end delay, packet loss and signal to interference plus noise ratio (SINR). In order to develop the simulation-based 5G mmWave evaluation framework, we have selected ns-3 as the network simulator and Simulation of Urban Mobility (SUMO) as the traffic simulator *(21, 22)*. The details have been presented in the following subsections.

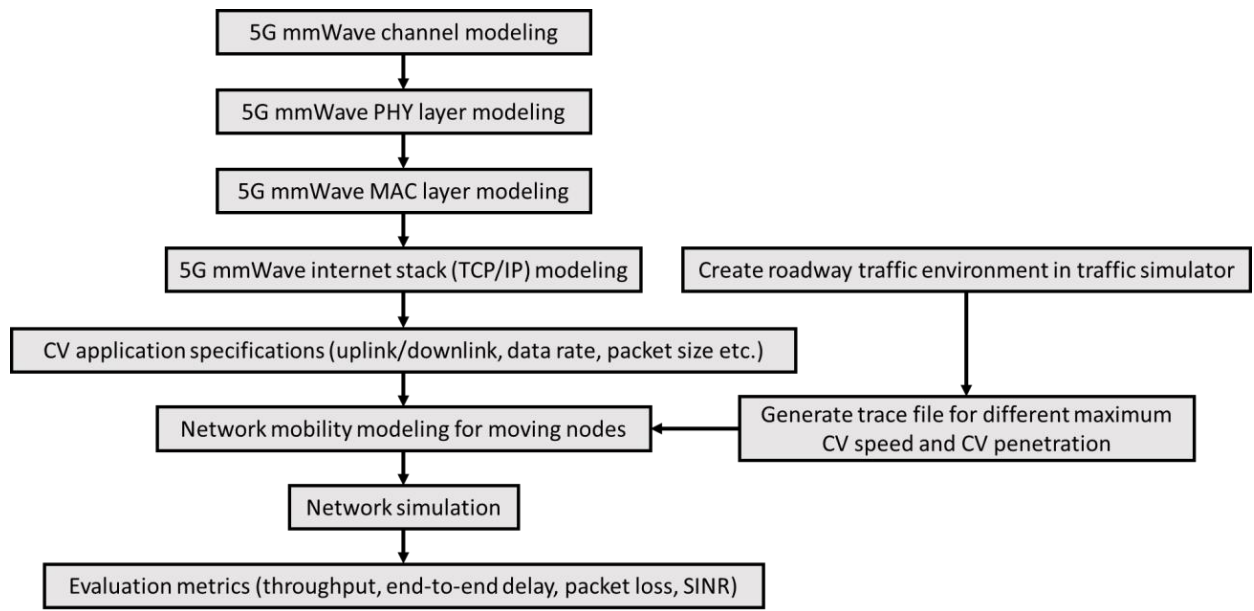

**Figure 2: Flowchart of 5G mmWave evaluation framework for CV applications**

**5G mmWave Modeling of Different OSI Layers**
To develop the evaluation framework of 5G mmWave for CV application, we have used the ns-3 network simulator *(21)*. Using ns-3, we model the communication between the CVs and the network infrastructure. To model the communication effectively, it is necessary to model different layers of the OSI model in a network device, starting from the physical layer to the application layer.

To simulate 5G mmWave communication in ns-3, we have used the mmWave module developed by New York University (NYU) Wireless and University of Padova *(23)*. The mmWave module offers a wide range of frequencies (i.e., 6~100 GHz) for simulating 5G communication. The 5G mmWave module architecture is built upon the ns-3 LTE module (LENA), which uses the evolved packet core (EPC) network for LTE communication *(24)*. All layers in the 5G mmWave module from the network layer and above use architecture of the LENA module. However, the mmWave PHY and MAC layers are designed specifically for mmWave. To simulate multiple CVs, we create multiple "MmWaveUeNetDevice" objects in ns-3, which represent the mmWave user equipment (UE) network devices. We also create one "MmWaveEnbNetDevice" object in ns-3 to simulate the radio stack in mmWave eNB *(23)*.

*Channel Modeling*
The first part of mmWave is channel modeling. We have used the NYU statistical model as the propagation path loss model in this study. This propagation model uses two separate equations for path loss considering line-of-sight (LOS) and non-line-of-sight (NLOS) scenarios *(23)*. Due to the computational burden of the simulation for having multiple UEs (CVs), we have used this as the propagation model because it reduces significant computation time. The communication channel matrices (i.e. beamforming vectors) are pre-generated and updated periodically (every 100 ms) to reduce computation burden during simulation while considering large-scale fading. The small-scale fading is calculated at each packet transmission. This model also considers blockages, as it overlays the statistical channel model with the blockage model and chooses the appropriate propagation path loss model. This model has only been calibrated for two frequencies, 28 GHz and 73 GHz. The 28 GHz spectrum is currently owned by mobile network operators, so it is not a free spectrum that can be used for vehicular communication *(11)*. Moreover, a higher frequency is desirable because it achieves higher throughputs and data rates, which is the focus of this study. The effect of blockages and mobility will be more severe in 73 GHz. Therefore, we have chosen 73 GHz as the carrier frequency in this study. The channel model also considers multipath interference by using an interference computation scheme that considers the beamforming vectors associated with each communication link. The beamforming vectors consist of several parameters, such as angle of departure and angle of arrival. The error model in mmWave module follows the LTE LENA error model *(24)*.

*PHY and MAC Layer Modeling*
Here, we describe the PHY and MAC layer properties of the mmWave module used in our simulation. The mmWave module uses a time division duplexing (TDD) frame and subframe structure following LTE but allows for more flexible allocation and placement of control and data channels within the subframe. It also uses the hybrid automatic repeat request (HARQ) based retransmission, which helps to do fast retransmission of data packets and increases the probability of successful decoding at the data receiving end. The HARQ model is combined with channel beamforming in order to ensure that all transmitted packets are received. In the MAC layer, time division multiple access (TDMA) is used as the default scheme for mmWave access because of analog beamforming. Analog beamforming refers to the transmitter antenna arrays aligning with the receiver antenna arrays to maximize the directional gain (rather than omnidirectional transmission). The adaptive modulation and coding (AMC) mechanism is used in our simulations, which uses the channel quality indicators (CQI) to update the modulation and coding. The MAC scheduler is based on a variable transmission time interval (TTI). We have used the round-robin scheduler in this study, which uses the orthogonal frequency division multiplexing (OFDM) and assigns OFDM symbols to flows in a round-robin order *(23)*.

*Internet Stack Modeling*
The remaining layers, including the internet stack (i.e. TCP/IP protocol suite), follow the LTE module. The "MmWaveHelper" object in ns-3 is used to model the 5G mmWave stack in simulation (e.g., channel, PHY, MAC). A UE (CV) is attached to the closest eNB at the start of the simulation. As we are interested in an end-to-end scenario, a packet gateway node (PGW node) is also created, which is connected to the backhaul LTE core network. The positions of the eNB are fixed, and the UE mobility model is derived from the TCL file generated from SUMO using "Ns2mobilityhelper" object. The final step is installing the applications in the CVs, eNB, and the remote host *(23)*. In this example, we are simulating downlink traffic, where the remote host sends data with a specific packet size at a certain bit rate, and the CVs receive the data while in motion. We have used UDP unicast transmission in this study since we want to achieve a higher bit rate at the cost of lower reliability.

*CV Application Specifications*
The application we have chosen for the performance evaluation is the high data rate-data download from roadside infrastructure to the CVs (i.e., downlink flow) which is an eMBB use case. The 5G mmWave was created with the promise of providing a high data rate for an high number of users. It is envisioned that one 5G mmWave base station can deliver 10 Gbps peak throughput, which is very useful when a high number of users are running multiple high data rate applications. Some examples of such applications for CVs are in-vehicle infotainment (i.e. high-definition video) and high-definition 3D map download. 5G mmWave communication utilizes higher frequencies (24~100 GHz) to achieve a high data rate but it is highly susceptible to blockages and mobility *(23)*. Our goal is to study the performance of 5G mmWave communication in a high data rate environment with multiple CVs. In this study, we will simulate multiple CVs running applications that need to download data at a high data rate, while moving at high speed through a corridor. We simulate different data rates (250 Kbps, 10 Mbps) and different packet sizes (1024 bytes, 256 bytes) in order to investigate its effect on communication performance. To show the improvement over the current state-of-the-art roadside network infrastructure, we will compare the performance with DSRC roadside unit (RSU).

**Roadway Traffic Environment Creation**
To evaluate the efficacy of 5G mmWave, we have selected one urban arterial from Greenville County, South Carolina, US, which is Woodruff Road. Over the timespan of 30 years (1960-1990), the surrounding area of the Woodruff Road has been transformed into a commercial hub from its early rural beginnings with the current average daily traffic being 40,000 vehicles. According to a survey conducted in 2017, among 4198 respondents, 67% and 81% identified the operational condition as 'very congested' for weekdays and weekends, respectively *(25)*. As a result, this corridor is very suitable for communication performance evaluation for CVs. We have used SUMO to simulate traffic flow through a corridor. To simulate a dense urban scenario, we have selected a portion of the woodruff road corridor and modeled it in SUMO. We generated different scenarios with varying numbers of CVs. The traffic on woodruff road varies by time of the day and our goal is to observe the communication performance while the traffic flow pattern changes in the corridor. Therefore, we have varied the CV generation rate (number of CVs through the corridor) and the speed of the CVs to see the effect on communication performance. By varying the number of CVs and speed of CVs, we generate multiple trace files. The trace file has information about each vehicle node, position, and speed, as well as the timestep when the vehicle data is captured. The trace files are used in the Network Simulator-3 (ns-3) to investigate the performance of 5G mmWave and DSRC wireless communications. **FIGURE 3** shows the deployment scenario for 5G mmWave in Woodruff road.

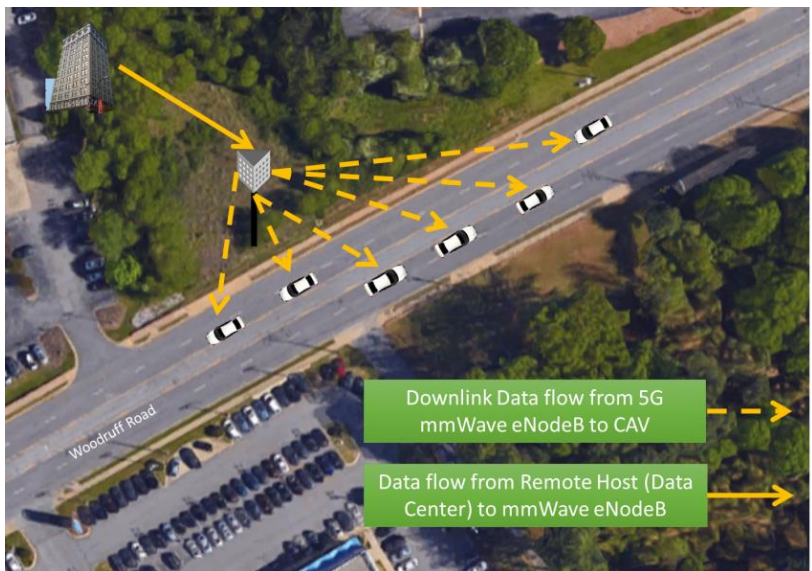

**FIGURE 3 CV application data flow (downlink) in Woodruff road using 5G mmWave eNB.**

**Base Case for Comparison: DSRC**
To model DSRC communication in ns-3, we have used the DSRC module already implemented in ns-3. The DSRC module or the wireless access in vehicular environment (WAVE) module in ns-3 is developed following the IEEE 802.11p, IEEE 1609 and SAE J2735 standards. As our scenario contains multiple CVs and one DSRC RSU, we have followed the vehicular ad-hoc network (VANET) example provided in ns-3 *(26)*. The propagation loss depends on two major factors, the distance between the communicating nodes and multipath fading. We have used the Friis propagation loss model to account for path loss due to distance and the Nakagami-m fast fading loss model to account for the path loss due to multipath fading. A transmission power of 20 dBm (equivalent to 0.1 Watts) is used. In terms of routing protocol, we have used the ad-hoc on-demand distance vector (AODV) routing protocol, which is the superior routing protocol for ad-hoc mobility scenarios *(26)*. The same trace files as the mmWave case are used for the node mobility. The RSU node is treated as a stationary node with a fixed position. A 5.9 GHz frequency is used as the spectrum of DSRC.

To compare with the 5G mmWave case, we install a similar application in the RSU and CVs. The RSU transmits data with a packet size of 1024 bytes and a bit rate of 250 Kbps. The CVs receive the packets if they are within the range of the RSU. The internet protocol (TCP/IP) is installed on the CVs and the RSU. UDP transmission protocol is used for sending the data packets. It should be mentioned here that we are using UDP unicast and not UDP broadcast, since the DSRC UDP unicast offers a higher maximum throughput compared DSRC UDP broadcast.

All the relevant parameters related to simulating 5G mmWave and DSRC in ns-3 are presented in **TABLE 1**. **FIGURE 4** shows the simulation scenario for DSRC.

**TABLE 1 Consideration for Wireless Communication for CVs**

| Wireless Communication Option | Considerations |
|---|---|
| **5G mmWave** | Frequency Band: 73 GHz<br>Packet size: 256, 1024 bytes<br>Data rate: 250 Kbps, 10 Mbps<br>Blockage Model: Enabled<br>Fading Model: Small-scale fading<br>PHY layer: Hybrid Automatic Repeat Request (HARQ) based retransmission<br>MAC Layer: Round Robin Scheduler |
| **DSRC** | Frequency Band: 5.9 GHz<br>Packet size: 1024 bytes<br>Data rate: 250 Kbps<br>Propagation Loss Model: Friis<br>Fast Fading Model: Nakagami-m<br>Routing Protocol: Ad hoc On-Demand Distance Vector (AODV) routing<br>Transmission and Receiving Gain: 20 dBm (0.1 Watts) |

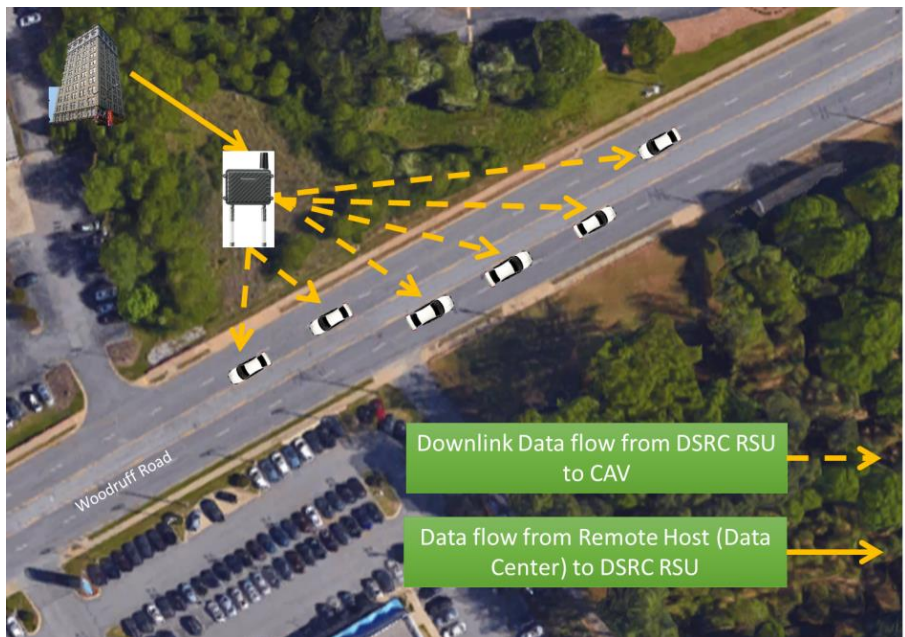

**FIGURE 4 CV application data flow (downlink) in Woodruff road using DSRC RSU**

**Performance Evolution metrics**
We have used the following evaluation metrics to evaluate both DSRC and 5G mmWave performance.
- Packet Loss Ratio (%) = $\frac{\sum_{i=1}^{n} L_i}{\sum_{i=1}^{n} S_i}$
  
  The ratio of the number of lost packets (L) for a receiver and the total number of packets (S) sent to the receiver. For a total time interval of *n* seconds, *i* is a particular simulation second.
- Mean End-to-End Communication Delay (ms) = $\frac{\sum_{i=1}^{n} D_i}{P_i}$
  
  Average delay (D) per received packet for a receiver where *P* is the total number of packets sent during the simulation run
- Receiver Bitrate or Throughput (Kbps) = $\frac{\sum_{i=1}^{n}(R_i*8)}{N}$
  
  Received packet bitrate, where R is the byte received per $i^{th}$ second, and N is the total simulation time
- Transmission Bitrate (Kbps) = $\frac{\sum_{i=1}^{n}(T_i*8)}{N}$
  
  Transmitted packet bitrate, with *T* is the byte transmitted per $i^{th}$ second, and *N* is the total simulation time
- Signal-to-interference-plus-noise ratio (SINR, db) = $\frac{M}{I+O}$
  
  *M* is the incoming signal's power, *I* is the interference of the other connected object, and *O* is the noise.

For the 5G mmWave scenario, the network simulator generates packet traces, which are post-processed to extract the packet information, along with timestamps (for the delay) and signal power (for SINR). For the DSRC scenario, we have used the flow monitor in ns-3, which monitors the packet flow between the communicating nodes *(27)*.

**DATA ANALYSIS**
We divide this section into multiple sub-sections. At first, we evaluate the performance of 5G mmWave for different CV penetration levels and CV maximum speeds. We also compare with DSRC performance to show the higher throughput capacity of 5G mmWave. Then, we investigate the effect of the packet size and application data rate on the performance of 5G mmWave network for CV applications.

**5G mmWave Performance for Different CV maximum speeds and Comparison with DSRC**
We keep the packet size (1024 bytes) and transmission (Tx) bit rate (i.e. data rate) (250 Kbps) fixed, because our goal is to investigate the effect of CV penetration levels and maximum CV speed on the performance of 5G mmWave network. A 250 Kbps downlink data flow is used for comparison with DSRC, since DSRC is limited on the maximum data rate it can provide.

We investigate the effect of maximum CV speed on the performance of 5G mmWave and DSRC. The number of CVs (i.e., 20 CVs) is fixed here. From Figure 5, we observe that 5G mmWave performance is not affected by the maximum CV speed for 20 CVs. The 5G mmWave eNB is able to provide average throughput of 200, 199, 201 Kbps (see Figure 5(a)), average packet loss of 21.8%, 22.3%, 21.5% (see Figure 5(b)), and average delay of 6.5, 3.8, 3.1 ms (see Figure 5(c)) for 35, 45 and 55 mph maximum speed of CVs, respectively. However, DSRC suffers from significant delay and packet loss due to the throughput requirements from multiple CVs (see Figures 5(b) and 5(d)), because DSRC has a maximum limit on the throughput for UDP unicast transmission. The DSRC RSU is able to provide average throughput of 18, 46, 14 Kbps (Figure 5(a)), average packet loss of 93.1%, 80.9%, 94.9% (see Figure 5(b)) and average delay of 2.8, 2.5, 3.6 s (see Figure 5(d)) for 35, 45 and 55 mph speed of CVs, respectively. From these results, it can be concluded that even for low data rates, the 5G mmWave is superior to the DSRC in terms of packet loss, throughput and delay. Moreover, the performance of 5G mmWave is not affected by the speed of the CVs.

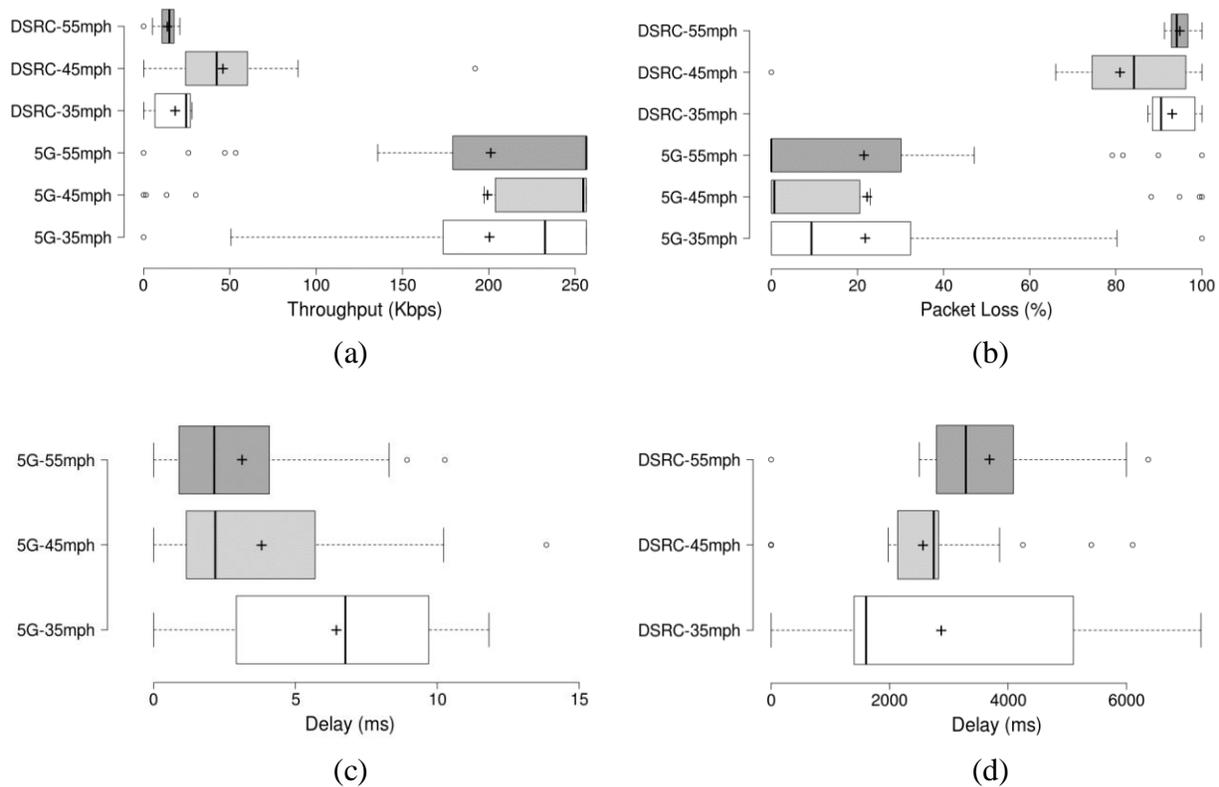

**FIGURE 5 Comparison between 5G mmWave and DSRC by varying the CV speed. (+ = Sample Mean)**

**5G mmWave Performance for Different CV Penetration Rates**
We investigate the effect of CV penetration rates on the communication performance. Here, we use two maximum speeds (35 mph and 55 mph) for CVs. The packet size is 1024 bytes and the Tx bit rate is 250 Kbps. From Figure 6, it can be observed that CV penetration level has noticeable impacts on the performance of 5G mmWave. We have performed simulations with 20 CVs and 40 CVs for both 35 mph and 55 mph. For both speeds, the increase in CV penetration level decreases the throughput and increases the delay and packet loss. At 35 mph maximum speed, the 5G mmWave eNB is able to provide average throughput of 200 Kbps and 163 Kbps (as shown in Figure 6(a)), average delay of 11.5 ms and 29.8 ms (see Figure 6(b)), and average packet loss of 21.9%, 36.4% (see Figure 6(c)) for 20 and 40 CVs, respectively. At 55 mph maximum speed, the 5G mmWave eNB is able to provide average throughput of 201, 141 Kbps (see Figure 6(a)), average delay of 8.1, 45.8 ms (see Figure 6(b)) and average packet loss of 21.5%, 44.7% (see

Figure 6(c)) for 20 and 40 CVs, respectively, at 55 mph speed. Here, we can observe that doubling the number of CV traffic reduces the throughput by a factor of 1.3, increases the average delay by a factor of 3.9 and increases the average packet loss by a factor of 1.9. The increase in the number of CVs increases the throughput requirements due to more applications downloading data using the 5G mmWave eNB. Moreover, more CVs mean a higher probability of NLOS scenarios, where any particular CV might be impeding the LOS for another CV. A scenario with higher number of CVs also creates more multipath interference, which results in a higher loss of packets and end-to-end delays. Another interesting finding is that the performance for 20 CVs remains unchanged for both speeds (35 and 55 mph), but for the 40 CV scenario, the performance of 5G mmWave eNB is worse when the maximum speed is higher. This suggests that at higher penetrations of CVs, a higher maximum speed of CV also becomes an issue for 5G mmWave communication.

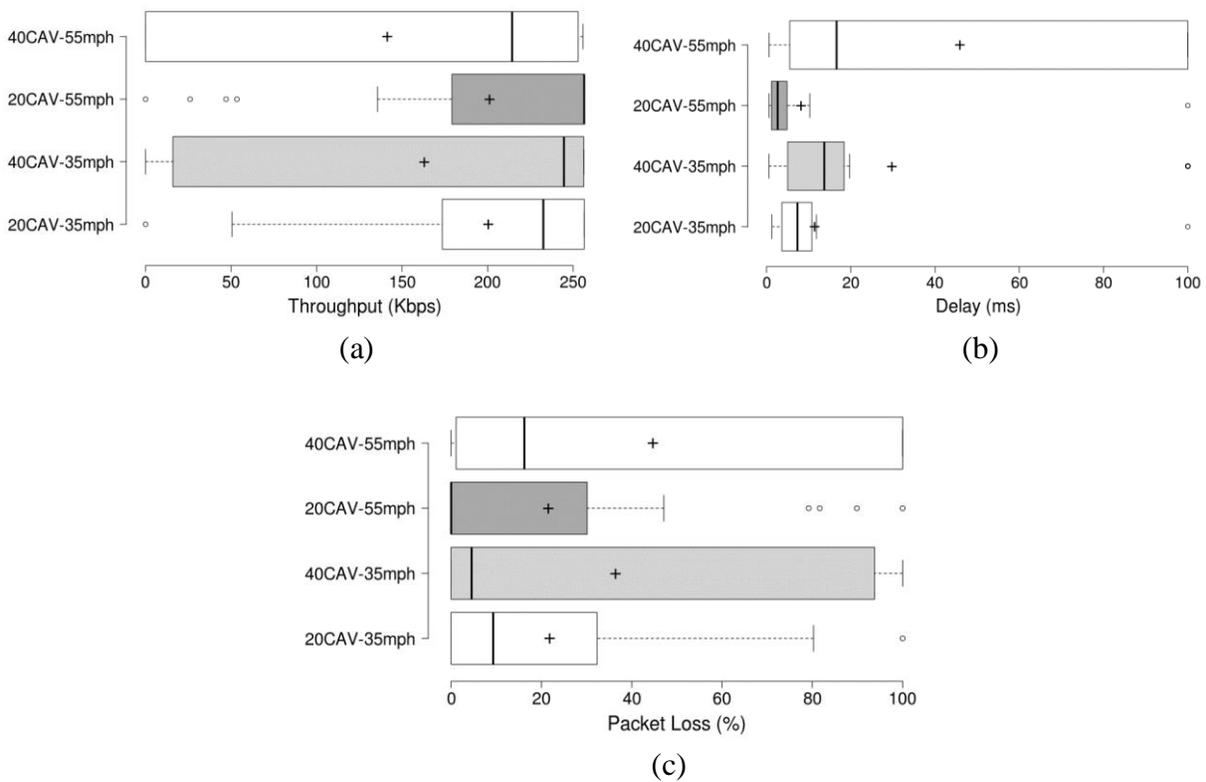

**FIGURE 6 5G mmWave performance variation for different CV penetration rates. (+ = Sample Mean)**

**5G mmWave Performance for Different Data Transmission Rates**

We examine the effect of a higher data rate on network performance of 5G mmWave. In previous results, we have assumed a packet size of 1024 bytes and a data rate of 250 Kbps for each CV application. However, we are interested in investigating the effect of higher data rates on the performance of 5G mmWave. Thus, we increase the data rate from 250 Kbps to 10 Mbps for each CV, keeping the packet size fixed at 1024 bytes. For this case, we use 20 CVs at an average speed of 45 mph. From Figure 7, we observe that a higher bit rate has significant impact on the network performance. The average throughput is higher for 10 Mbps case because the Tx bit rate is higher, but we can perform a one-to-one comparison in terms of delay and packet loss. We also observe that for 10 Mbps case, the average delay is 59.9 ms compared to the 8.8 ms for the 250 Kbps case, so the delay increases by a factor of 6.8 when the bit rate increases by a factor of 40 (see Figure 7(a)). The average packet loss increases from 22.3% (for the 250 Kbps case) to 63.3% (for the 10 Mbps case), which is an increase by a factor of 2.8 (see Figure 7(b)).

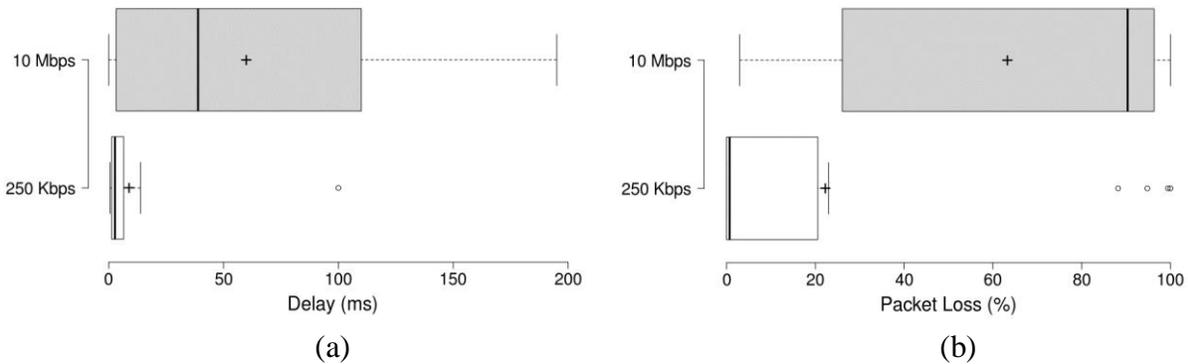

(a)                  (b)

**FIGURE 7 5G mmWave performance variation with data transmission rates.**
**(+ = Sample Mean)**

**5G mmWave Performance for Different Packet Sizes**
Now, we examine the effect of packet size on network performance of 5G mmWave. For the previous results, we have assumed a packet size of 1024 bytes and a Tx bit rate of 250 Kbps for each CV application. In this scenario, we decrease the packet size from 1024 bytes to 256 bytes for each CV, keeping the data rate fixed at 250 Kbps. For this case, we use 20 CVs at an average speed of 45 mph. From Figure 8, we observe that a change in packet size has no significant impact on the delay (Figure 8(a)) and packet loss (Figure 8(b)). The average delay is 9.8 ms compared to the 8.8 ms for the 1024 bytes case, and the average packet loss is 21% compared to 22.3% for the 1024 bytes case.

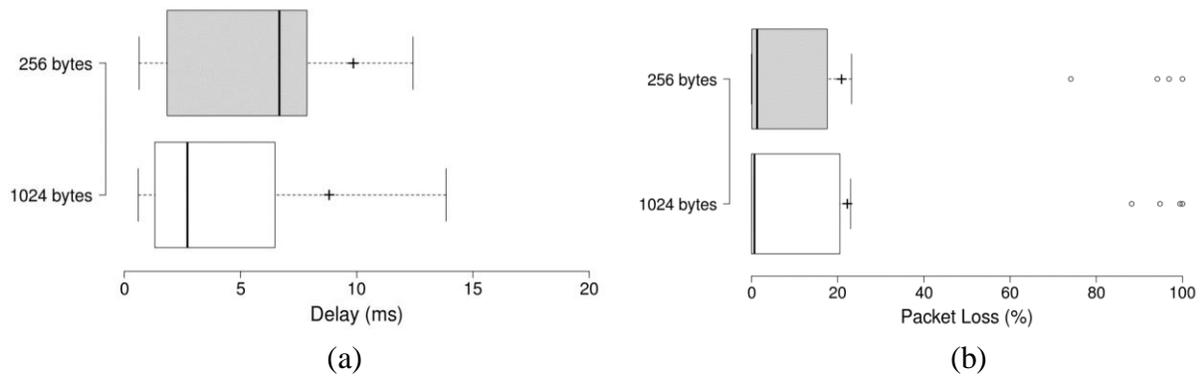

(a)                  (b)

**FIGURE 8 5G mmWave performance variation with packet size**
**(+ = Sample Mean)**

**5G mmWave Performance Evaluation using SINR**
We have also measured the quality of 5G communication in terms of SINR. SINR is a wireless quality indicator where a higher value represents a higher quality of wireless communication with higher throughput. Figure 8 shows the SINR distribution of different evaluation scenarios. From Figures 9(a) and 9(b), it is clear that interference and noise increases with the higher number of CVs. For CV=20, the SINR is between -20 dB and 0 dB, but for CV=40, the SINR is spread out in the range of -35 dB and 0 dB. In Figure 9(c), the SINR distribution is improved compared to Figures 9(a) and 9(b) due to the higher Tx bit rate and higher throughput. Throughput is logarithmically proportional to SINR, as defined by the Shannon-Hartley theorem *(28)*.

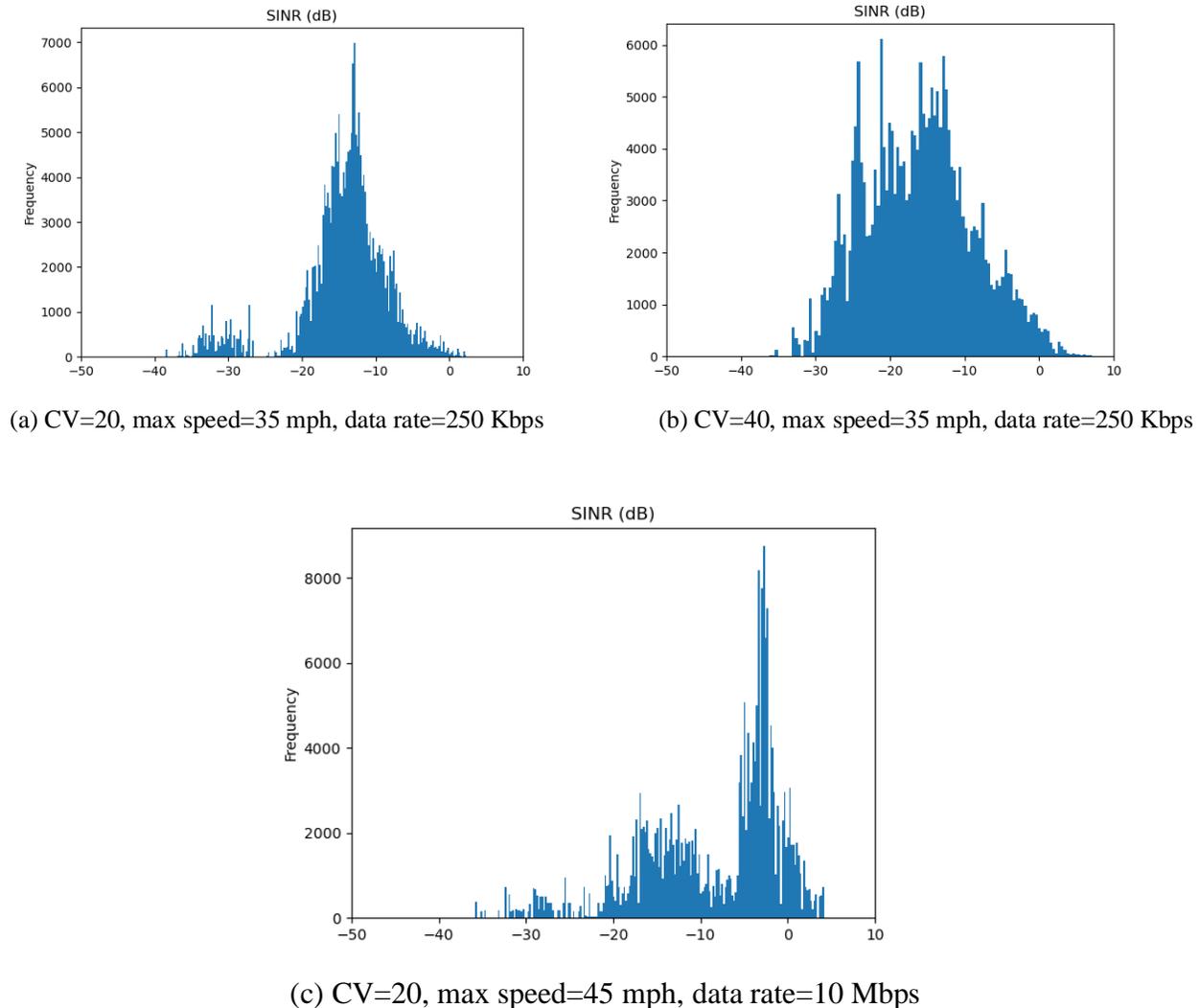

(a) CV=20, max speed=35 mph, data rate=250 Kbps  (b) CV=40, max speed=35 mph, data rate=250 Kbps

(c) CV=20, max speed=45 mph, data rate=10 Mbps
**FIGURE 9 5G mmWave SINR distribution for different scenarios**

**CONCLUSIONS**
CVs need reliable and fast communication with a high data rate to support different CV applications. Meeting all of these requirements has been a major challenge for existing widely deployed wireless communication services, such as 4G, LTE and LTE-V. The 5G mmWave communication is a promising wireless communication option to support CV applications, especially in a dense urban area. The findings from this study establish that 5G mmWave can be the enabler of CVs in dense urban areas. Using the evaluation framework developed in this paper, agencies can evaluate 5G mmWave to support CVs in urban areas containing heavy pedestrian traffic like downtown and commercial zones. However, the 5G mmWave faces many challenges in the dynamic connected vehicle scenarios. Although the speed of CVs does not affect the performance too much, the increase in CV penetration levels will have a major impact on communication throughput, delay and packet loss. Moreover, the speed of CVs also becomes an issue when the number of CVs are high. Finally, the increasing data rates required by CV applications pose a major problem for 5G mmWave. Although 5G mmWave is showing promising performance, its signal is susceptible to blockage and non-line-of-sight, which is why the base stations have to be closely spaced. Having closely spaced mmWave base stations is an ideal scenario for an urban environment where CVs will be operated by communicating with other CVs, connected pedestrians, and other external data sources. More research is needed in this domain, including innovative beamforming, beam selection, and beam tracking techniques to improve the communication efficiency. Additionally, testing is required on the performance of a heterogeneous network with 5G mmWave and other communication technology to provide a reliable communication system for a long travel area, such as intercity highways, where CVs are switching between communication technologies based on their availability and coverage in the travel area.

## AUTHOR CONTRIBUTION STATEMENT
The authors confirm contribution to the paper as follows: study conception and design, Zadid Khan, Sakib Mahmud Khan, Mashrur Chowdhury; data collection, Zadid Khan, Sakib Mahmud Khan; interpretation of results, Zadid Khan, Sakib Mahmud Khan, Mashrur Chowdhury, Mizanur Rahman, Mhafuzul Islam; draft manuscript preparation, Zadid Khan, Sakib Mahmud Khan, Mizanur Rahman, Mhafuzul Islam, Mashrur Chowdhury. All authors reviewed the results and approved the final version of the manuscript.

## ACKNOWLEDGEMENT
This work is based upon the work supported by the Center for Connected Multimodal Mobility ($C^2M^2$) (a U.S. Department of Transportation Tier 1 University Transportation Center) headquartered at Clemson University, Clemson, South Carolina, USA. Any opinions, findings, conclusions and recommendations expressed in this material are those of the author(s) and do not necessarily reflect the views of $C^2M^2$, and the U.S. Government assumes no liability for the contents or use thereof.